# Parameters Differentiating the Characteristics and Security of Military Information Systems

Muhammad Farooq-i-Azam
COMSATS Institute of Information Technology
Lahore, Pakistan

farooq@chase.org.pk

Dr. Muhammad Naeem Ayyaz
University of Engineering and Technology
Lahore, Pakistan

mna@uet.edu.pk

## ABSTRACT
Revolution in the area of information technology has brought about changes in many spheres of life. Today, information systems are being used in very sensitive areas such as defence and missile control systems, nuclear plants, etc.  Not only has it changed how business is conducted, it has also brought about entirely new paradigms like that of information and cyber warfare. Similarly, one of the many impacts that it has made, is how wars are fought. For all what it has contributed, the information stored on digital devices and computers has become a precious resource and special measures are taken to guard it against attacks from malicious users. These special measures are needed by any enterprise be it a business firm, a commercial entity, a government agency or a military organization. However, requirements and specifications for information security and assurance for a military organization are essentially different from those of commercial or business applications. This paper highlights and discusses various aspects related to the security of information resources of military importance and outlines certain parameters that should be taken into consideration when talking about the security of military information resources. W also describe the role satellite reconnaissance can play in cyber war. Authors proclaim that this paper is first such attempt to correlate cyber war with satellite reconnaissance.

## 1. INTRODUCTION
Information systems in current age use digital processing and storage. Like other spheres of life military is also increasingly dependent upon digital information systems. Digital systems and networks because of their inherent characteristics are susceptible to various kinds of remote attacks. Therefore, the military needs to adopt special measures to protect the confidentiality, integrity and availability of its own sensitive information resource related to its own functioning and operations. Furthermore, when attacks against the information resources of other important non-military institutions of a country are mounted in an organized manner by a hostile country, the role of military again comes into play which is to protect and defend the information resources of its country and to launch offensive attacks against the perpetrator as a tactical and strategic measure. Recently, information and cyber warfare has been focus of much attention around the world. In particular, the CIA, intelligence outfit of USA, carried out cyber warfare exercises in 2005 [8]. It is also important to mention that in May 2010 US army has announced the creation of a cyber war command and appointed a four star general to lead it [4]. This significant step forecasts the role of information technology in future wars and also nature of these wars. Other militaries will sooner or later follow these same steps.

For the information and cyber warfare, the cyber arms needed are peculiar information systems particularly designed and developed for this purpose. At the heart of these systems are software and tools both for attack and defence. Traditionally, there are a lot many such tools available including free and open source developed by hackers and security professionals and commercial software and tools developed by companies. However, these traditional tools though suited to attack or defend commercial and business systems, are not suited to the needs of a military.

One of the major reasons for this is the simple fact that software tools developed by someone else cannot be relied upon for mission critical tasks. There may be intentional or un-intentional trap doors left in such tools and software so that it may be rendered inoperable at a critical moment.

Another important factor is the fact that the enemy will not be using traditional tools and software for the reasons stated above. Therefore, capability, capacity and characteristics of such systems are not known in advance. As a result, the countermeasures required against such attacks cannot be based upon traditional techniques and rather are tailored to use by the military.  For example, to communicate some information related to trade and commerce, one may rely on openly available commercial solutions and encryption algorithms.  However, information which is related to defence and military operations needs to be taken more seriously. Instead of deploying tools and software using a third party encryption algorithm, the military would need to develop and implement its own encryption algorithm and software for this purpose. In addition, like an integrated circuit has different military specifications, this software for use by the military would also have different specifications. By the same token, the attack mechanisms and countermeasures used by the military for information systems are different from traditional systems.

Furthermore, the military information systems may not simply comprise of merely computer systems, computer networks and software tools. For example, information and cyber war will inevitably also use military reconnaissance satellites. Therefore, implications and effects of such satellite systems in a cyber war should also be taken into account and capacity to counter such a scenario should also be developed.

It is to be further noted that military information systems is not an open topic of research due to obvious reasons and not much information is available in this area. The research presented in this paper is first-hand research and is derived from authors' experience and consistent observation of military use of the Internet over the years.

## 2. MILITARY INFORMATION SYSTEMS REQUIREMENTS

We have stated previously the reasons for military information systems to have different characteristics and security requirements. In this section, we give brief details of such requirements and characteristics regardless how easy or difficult these requirements are as far as their implementation is concerned.

### 2.1 Anonymity

In the case of commercial and business enterprises, the IP address blocks used by the organization may be known to the public and probably cannot be hidden from being discovered. There are various means through which an attacker can enumerate the IP address blocks of such organizations.

However, in the case of military organizations, there are various levels of security and confidentiality that are ascribed to different institutions of the military. In this case, the IP address blocks used by certain unclassified institutes may be known to the public. However, IP blocks used by certain other more classified projects should be hidden from public view. By stating that an IP address block is in public knowledge, we mean, for example, that:

1- Geographic location of an IP address i.e. country, city can be found out.
2- Organization to which the IP address is assigned can be found out.
3- How do Internet registrars respond to queries against these IP address blocks.

In the case of information warfare with military implications, it is very important that the source network used for the attack cannot be traced back to. Therefore any design and audit of such a military network needs special considerations which are different from those of commercial and business networks. It is to be noted that how such a level of anonymity may be achieved is entirely another area of research. In brief, the level of anonymity one can achieve is relative to the capability of the enemy.

### 2.2 Traffic Concealment

It is usual in military to keep a tab on the activity of the enemy. Translated to the cyber warfare, this means that a military should keep a tab on the defence networks of the hostile countries and similarly expect the same from the other side and adopt means to hide or conceal its networks activity and traffic. If the enemy cannot penetrate a network, it would simply be interested to know what kind of traffic is flowing to and from network perimeter. What kinds of search queries are flowing to and from the network, the email messages, who are the people or networks at the other end, etc. The exact solution for this problem is encryption. However, not all destination networks with which a source network communicates support encryption. Therefore, a means should be found out to conceal the traffic of a defence network. It is to be noted that requirement of concealment of traffic is another layer of security on top of anonymity. Appropriate levels of traffic concealment may be applied to different computer systems and networks having various levels of security.

### 2.3 Specialized Encryption

Conventional encryption algorithms and those available commercially are not suitable for the purposes of defence and military purposes. On the other hand, a military would never trust or share the encryption algorithm it uses itself with another country. Or if does, then the communication that uses it is secure as long as the provider does not decide to just intercept and decipher the cipher text. In other words, an encryption algorithm is not something that you can buy from somewhere. For a trusted information and communication system, the military needs to build, develop and deploy its own encryption algorithms. It is general knowledge that the countries like US do not even allow the export of commercial versions of certain encryption algorithms beyond certain key lengths.

### 2.4 Classification of Computer Systems

The computer systems, network and communication hardware employed by the military is usually graded and assigned different security levels. Before the equipment is deployed, it undergoes various levels of security evaluations and checks. As a very crude example, the computer systems may be assigned different security levels ranging, say, from 1 to 7 with 1 being assigned to a non-classified system and 7 to a highly classified system. Different evaluation criteria may be developed for these systems which again may range from checks on the design and various stages of development of the equipment and then final testing before it is deployed. It is to be noted that these checks are employed not only on the software but also on the firmware and hardware i.e. evaluation starts not on procurement but much earlier at the design stage of the equipment. For militaries which do not design and develop their equipment, this step is skipped thereby compromising the security of these systems. If a military cannot follow the steps described, it is always better to know where it stands if the precautions described are not followed.

It is pertinent to mention here that hardware Trojan horses at the level of circuit board and even integrated circuits are possible and is focus of current research [11]. Similarly, BIOS rootkits have also been designed and hence any hardware piece of equipment is susceptible to almost same kinds of attacks as software systems.

#### 2.4.1 Hardware Trojan Horses

Consider the case of a stand-alone digital system i.e. a digital system that is not part of any network. As the digital device does not interact with any external network, it may be thought that no attacks can be mounted against the device. However, it is still possible for a malicious design engineer to leave a malignant hole i.e. a Trojan horse in the system. For example, a design engineer could program a peripheral device to run correctly for all operations except, say, #2600th, or program it in such a way to behave erratically after certain number of operations or under a certain critical condition.

The designer could also leave a hardware Trojan horse that can be controlled remotely, possibly using a radio channel, in an otherwise standalone digital system. It may be noted that the Trojan horse may be part of a digital system in the shape of discrete components or at the level of transistors in an integrated circuit. In the first case, the Trojan horse may be present on a circuit board of the system and in the second case, it may be present inside an integrated circuit in the shape of transistors.

Above descriptions may seem unlikely to some, and therefore it would be interesting to cite just two instances where hardware Trojan horses were detected and found spying on unauthorized information. In the first instance, Seagate external hard drives were found to have a hardware Trojan horse that transmitted user information to a remote entity [10]. In the second instance, Prevelakis et al. report in [11] that, in 2006, Vodafone routers installed in Greece were altered in such a manner so as to allow eavesdrop phone conversations of the prime minister and many other officials.

The hardware Trojan horse may be implanted in the hardware system at any of the various stages from design to implementation including plantation. This includes the possibility of implant by the designer at the level of behavioral description, by a third party synthesis tool or by the fabrication facility.

## 2.5 Software Engineering

Even in the world of hackers (we use the term in positive sense), it is considered bad practice to use someone else's tools or software to test or penetrate a remote system and such folks are somewhat derogatorily termed as script kiddies. Almost always, hackers leave a trapdoor or some sort of bug in the "tool" they distribute to others. For example, even Linux distributions like Slackware and Debian which are otherwise considered very secure, leave a few unnecessary remote services running which may allow an attacker to intrude into the system. Other times, hackers may purposely develop a tool such that it leaves a trace or some sort of signature so that its action could later be tracked. Therefore, it is imperative for the military to develop its own set of cyber arms i.e. software and tools needed for military operations. As the only resource required for the development of such software is human resource, the cost factor is not important. Indeed, the cost of self-built software may be quite lower than an equal quality software purchased from a third party with the added advantage of technology transfer.

## 2.6 Satellite Reconnaissance

From the current commercial satellites, the GeoEye-1 launched in 2008 and operated by the private firm GeoEye located in the US has a stated resolution of 41 cm or 16 inches which is the highest amongst commercial satellites [7]. Furthermore, just after two years i.e. in 2012, the same company GeoEye would launch satellite GeoEye-2 that would have resolution twice higher resolution i.e. 20.5 cm or 8 inches. We can infer that commercial satellite resolution increases at the rate of 2 every four years. Rate of increase of resolution of military and reconnaissance satellites must be higher than this as primary research in this area is carried out by military. However, let us stick with the conservative rate of increase of resolution as 2 per four years. Further, resolution of military and reconnaissance satellites can be estimated by comparing the size of corresponding commercial and military organization that run this business When this comparison is made, the private firm GeoEye comes out to be much smaller enterprise than their military counterparts i.e. National Reconnaissance Office (NRO), the organization which is responsible to build and operate US spy satellites. For example, if we consider the annual financial budget figures, NRO has annual budget of almost US$ 8 billion whereas GeoEye has total assets worth almost US$ 790 million i.e. less than one billion. In terms of number of employees, GeoEye has only 410 employees as compared to almost 3000 people employed by NRO. Therefore, we can safely assume that NRO satellites have a resolution that is at least 10 times better than that of commercial satellites. Therefore, corresponding to GeoEye-1, a NRO military satellite would have an estimated resolution of 4.1 cm or 1.6 inches i.e. it can take photograph of an individual. And just after two years i.e. by 2012, it would have an estimated resolution of 2.05 cm or 0.8 inches i.e. it would be able to what is being typed on a keyboard.

This estimate is corroborated from other sources as well. Dwayne A. Day in [6] states that US spy satellite KH-8 could be used to take photographs of smallest objects and even portraits of individuals. We quote Dwayne A. Day from his article [6] as:

*"The KH-8 could apparently see objects on the ground as small as a baseball"*

*"There was the time that some Air Force officers used one to take a self-portrait".*

It is to be noted that the spy satellite KH-8 for which above remarks are made was operated in early 1970s. From this, we can once again infer the resolution of current spy satellites. If we trust above remarks and also consider the fact that it was almost 30 years ago that KH-8 was operated, it is certain that resolution of the current spy satellites should rather be better than our estimates.

Please note further that satellites have a variety of remote sensing capabilities which may include seeing through building structures as well. This may well be derived from the fact that satellites have long been used to locate minerals and structures underground.

It is to be noted that, resolution of reconnaissance satellites is not stated publicly due to obvious reasons and only estimates can be made. Resolution of these satellites as calculated above in this paper is what the authors consider as the best estimate.

Consider above stated capabilities of spy and reconnaissance satellites being available for cyber and information warfare. Once the geographical location and name of the organization where the

IP address of an enemy is located is found out, the spy satellite can look directly at the person and the systems being used by them. Obviously, this sort of capability would have a decisive effect on the outcome of the war.

## 3. FUTURE WORK

Each of the characteristics of a military information system that we described in this paper leads to a different and its own area of research. The attributes and dimensions of a military information system that we have provided here just give an abstract and upper layer description which are by no means complete and merely provide a starting point in this direction. Each of these areas needs further investigation so as to provide lower level details of a military information system.

## 4. CONCLUSION

In this paper, we have described and presented parameters that differentiate a military information system from commercial and business systems. In short, military information systems have requirements which are essentially different from those of commercial and business systems simply because of the reason that military and business systems are used for entirely different purposes. Some of such parameters are specialized encryption, anonymity, traffic concealment, assignment of security levels to computer systems according to their roles, etc. We have also suggested and discussed the role that spy satellites can play in a cyber war.